# Ring Polymer Molecular Dynamics in Gas-Surface Reactions: Inclusion of Quantum Effects Made Simple

Qinghua Liu[1], Liang Zhang[1], Yongle Li[2,*], and Bin Jiang[1,*]

[1]*Hefei National Laboratory for Physical Science at the Microscale, Department of Chemical Physics, Key Laboratory of Surface and Interface Chemistry and Energy Catalysis of Anhui Higher Education Institutes, University of Science and Technology of China, Hefei, Anhui 230026, China*

[2] *Department of Physics, International Center of Quantum and Molecular Structures and Shanghai Key Laboratory of High Temperature Superconductors, Shanghai University, Shanghai 200444, China.*

*: corresponding authors: yongleli@shu.edu.cn, bjiangch@ustc.edu.cn

**Abstract**

Accurately modeling gas-surface collision dynamics presents a great challenge for theory, especially in the low energy (or temperature) regime where quantum effects are important. Here, a path integral based non-equilibrium ring polymer molecular dynamics (NE-RPMD) approach is adapted to calculate dissociative initial sticking probabilities ($S_0$) of $H_2$ on Cu(111) and $D_2O$ on Ni(111), revealing distinct quantum nature in the two benchmark surface reactions. NE-RPMD successfully captures quantum tunneling in $H_2$ dissociation at very low energies, where the quasi-classical trajectory (QCT) method suddenly fails. Additionally, QCT substantially overestimates $S_0$ of $D_2O$ due to severe zero point energy (ZPE) leakage, even at collision energies higher than the ZPE-corrected barrier. Instead, NE-RPMD predicts $S_0$ values of $D_2O$ in much improved agreement with reference results obtained by the quantum wavepacket method with reasonable corrections of the thermal contribution. Our results suggest NE-RPMD as a promising approach to model quantum effects in gas-surface reactions.

TOC graphic

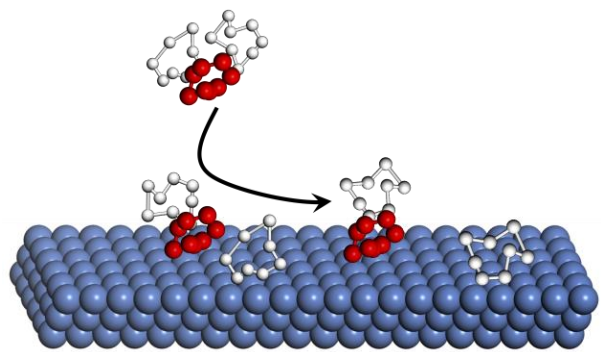

Dissociative adsorption of small molecules at surfaces is the initial and often rate-limiting step in many interfacial processes such as heterogeneous catalysis and corrosion. Initial sticking probability ($S_0$) is an important observable to reveal the adsorption mechanism and dynamics at solid surfaces, which can be now precisely measured as a function of incidence energy by molecular beam experiments[1-2]. $S_0$ can be $10^{-5}$ or even lower at low energies in many highly activated systems, $e.g.$ methane and water dissociative chemisorption on metal surfaces[1, 3], representing indispensable steps in methane steaming reforming. Given the large number of degrees of freedom (DOFs), however, it is very challenging to accurately predict $S_0$ from first-principles calculations.

Ideally, given an accurate global potential energy surface (PES), exact quantum initial sticking probabilities can be extracted by fully-coupled quantum dynamical (QD) methods, including both time-dependent[4-6] and time-independent[7] implementations. While such molecule-surface PESs can be now routinely developed[8-9], high-dimensional QD calculations remain extremely difficult and are limited to involve at most nine molecular DOFs so far[10], due to their poor scaling with dimensionality. Alternatively, the quasi-classical trajectory (QCT) method is much more efficient to model surface reaction dynamics and visualize the associated mechanism[4]. In QCT calculations, the zero point energy (ZPE) of the reactant is approximately included, while the atomic motion is evolved classically. Such QCT applications in $H_2$ activated dissociation on various metal surfaces do reproduce the QD calculated $S_0$ values above the ZPE-corrected barrier quite well[4, 11]. When explicit PESs are unavailable, ab initial molecular dynamics (AIMD) simulations, in which the energies and forces are calculated on-the-fly, have also been performed to study surface reactions[12]. Although AIMD is computationally expensive, it incorporates both molecular and surface DOFs in a full-dimensional fashion, enabling a reliable, and sometimes chemically accurate description of surface reactions involving polyatomic molecules (*e.g.* CHD$_3$)[13-14].

However, QCT and the quasi-quantized version of AIMD intrinsically suffer from the artificial ZPE leakage and neglect quantum tunneling, owing to their classical nature in the propagation. Indeed, when the incidence energy is lower than the ZPE corrected barrier height, it was found that the large amount of ZPE in polyatomic molecules like methane may improperly flow into other DOFs, leading to a significant overestimation of total reactivity and scrambling the vibrational mode specificity in methane dissociation on metals[15-16]. Such an overestimation of $S_0$ by QCT has also been seen in the non-activated H$_2$+Pd(100) system[17]. On the other hand, in H$_2$ dissociation on Pd(111), the vibrational energy is found to be unphysically transferred to the motion normal to the surface, resulting in an unexpected decrease of trapping meditated reaction probability at low energies[18]. Various efforts have been devoted to alleviating the ZPE problem in QCT simulations for different systems[19-22]. A common strategy in bimolecular reactions to discard trajectories violating the ZPE of products[23], is not applicable in molecular dissociation on a surface, in which the dissociated co-adsorbates are typically not well separated. Other strategies are more complicated and their effectiveness is often system dependent.

Recently, ring polymer molecular dynamics (RPMD), an approximate version of path integral molecular dynamics (PIMD), has emerged as an efficient approach that can mimic quantum dynamics in complex systems with an affordable cost[24]. In particular, with the ansatz of quantum-classical correspondence proposed by Craig and Manolopoulos[25-26], RPMD has been successfully applied in calculating quantum rate constants in a variety of gas phase reactions[27-28] and hydrogen diffusion on metal surfaces[29]. More recently, Miller and coworkers showed that RMPD could be also used in non-equilibrium conditions[30], for example, in the H atom scattering on graphene[31]. Suleimanov *et al.* have performed RPMD simulations for studying direct dynamics of ion-molecule reactions and obtained thermal rates and product branching ratios[32]. These studies motivate us to extend the RPMD approach to molecule-surface scattering problems. In particular, we report here for the first time calculations of $S_0$ down to the very low energy regime where quantum effects may play a dominant role.

Since there have been excellent reviews on the RPMD theory[24, 28], we only provide a brief summary here. This method is based on an isomorphism between the path integral representation of the quantum mechanical partition function and the classical counterpart of a fictitious ring polymer[33]. With this isomorphism, quantum mechanical partition function is mapped onto the classical counterpart ($Z^P$) given by[25],

$$Z^P = \frac{1}{(2\pi\hbar)^P} \int d\mathbf{p} \int d\mathbf{q} \ e^{-\beta_P H_P(\mathbf{p},\mathbf{q})},\tag{1.1}$$

where $\mathbf{p}$ and $\mathbf{q}$ are the collections of the momentum and position vectors of the system with $P$ replicas, $\beta_P = \beta/P$ , and $\beta = 1/(kT)$ with $k$ being the Boltzmann constant. The isomorphic ring polymer Hamiltonian $H_P(\mathbf{p},\mathbf{q})$ consisting of $N$ atoms is extracted as[25],

$$H_P(\mathbf{p},\mathbf{q}) = \sum_{k=1}^{P} \sum_{j=1}^{N} \left[ \frac{(\mathbf{p}_j^k)^2}{2m_j} + \frac{1}{2} m_j \omega_P^2 (\mathbf{q}_j^k - \mathbf{q}_j^{k-1})^2 \right] + \sum_{k=1}^{P} V(\mathbf{q}_1^k, \mathbf{q}_2^k, ..., \mathbf{q}_N^k),\tag{1.2}$$

where $m_j$ is the $j$th atomic mass that is used for each bead of this atom, $\omega_P = 1/\beta_P$ is the frequency of the inter-bead harmonic potential, $\mathbf{q}_j^k$ and $\mathbf{p}_j^k$ represent the position and momentum vectors of the $k$th bead of the $j$th atom with $\mathbf{q}_j^0 = \mathbf{q}_j^P$, and $V$ is the molecular interaction PES. Replacing each atom by this $P$-bead ring polymer, RPMD approximates the quantum time evaluation using classical-like trajectory propagation in an extended phase space. Apparently, RPMD collapses to the classical MD for a single bead ($P$=1). As $P$ goes to infinite, on the other hand, RPMD time correlation function becomes exact in the short time, high temperature, harmonic oscillator, as well as parabolic barrier limits[24]. RPMD naturally incorporates the quantum ZPE into the system, which preserves the quantum Boltzmann distribution at a given temperature and thus avoids unphysical ZPE leakage by construction[34]. In addition, RPMD captures tunneling as well, even at very low temperature in the deep quantum tunneling regime[35]. This has been discussed by Richardson and Althorpe via an insightful connection between RPMD and semiclassical instanton theory[35].

For describing the gas-surface collision process with a specific incidence energy, we use the non-equilibrium RPMD (NE-RPMD) Hamiltonian[30],

$$H'_P(\mathbf{p}, \mathbf{q}) = \sum_{k=1}^{P} \sum_{j=1}^{N} \left[ \frac{(\mathbf{p}_j^k - \Delta \mathbf{p})^2}{2m_j} + \frac{1}{2} m_j \omega_P^2 (\mathbf{q}_j^k - \mathbf{q}_j^{k-1})^2 \right] + \sum_{k=1}^{P} V(\mathbf{q}_1^k, \mathbf{q}_2^k, ..., \mathbf{q}_N^k) \qquad (1.3)$$

where an initial momentum impulse $\Delta \mathbf{p}$ is added that corresponds to the initial translational energy defined in molecular beam experiments. Miller and coworkers[30] have shown that many desirable features of equilibrium RPMD in various limits are preserved in non-equilibrium conditions and RPMD is able to achieve similar accuracy for calculating equilibrium and non-equilibrium correlation functions.

Our NE-RPMD simulations of gas-surface reactions are implemented in two stages, following the recent work of Suleimanov and coworker[32]. An equilibrium PIMD run with a thermostat is first performed for the reactants in the asymptote (here the molecule only which is far above the fixed surface) at a given temperature. In this stage, the center of mass translational energy is removed so that the internal energy contains only the internal contribution subject to the quantum Boltzmann distribution. The positions and momenta of ring polymers are taken from random snapshots after equilibration, which ensures subsequent NVE trajectories with correct quantum internal energies that explore different regions in configurational space[36]. The collision upon surface is driven by a momentum impulse acting on the molecule along surface normal with no thermostat, corresponding to the specific translational (or normal incidence) energy in molecular beam experiments. With these well-defined initial conditions, the NE-RPMD reaction probability can be obtained in the similar way as that in QCT, except that the atomic position is replaced by the centroid. More technical details are discussed in the Supporting Information (SI).

Despite its great success in predicting quantum thermal rate constants at low temperatures[28], RPMD has never been tested in calculating the diminishingly small reaction probabilities at very low translational energies for an activated chemical reaction. As a proof of concept, this is validated here in two benchmark systems, namely $H_2$ dissociation on Cu(111) and water dissociation on Ni(111). These two reactions represent the simplest diatomic and polyatomic reactions on metal surfaces, for which the very accurate QD $S_0$ are available within the Born-Oppenheimer static surface (BOSS) approximation, enabling a quantitative comparison among RPMD, QCT, and QD results. As our purpose is to validate the RPMD approach, we choose an analytical and efficient London-Eyring-Polanyi-Sato PES parameterized by Dai and Zhang (DZ) based on limited density functional theory (DFT) data[37]. The DZ PES has been found to describe the ground state reactivity[38] and the orientational effects depending on the translational energy[39] reasonably well. Although a chemically accurate PES was also reported (though not publicly available)[40], it was based on interpolation and less efficient for running numerous RPMD trajectories at very low energies. For $D_2O+Ni(111)$, we take a nine-dimensional neural network PES accurately fitted to over twenty thousand DFT data[8], which naturally preserves permutation and surface symmetry and is therefore suitable for large scale MD simulations. We select $D_2O$ instead of $H_2O$ because there were some previous QCT[8] and QD[41] data available for straightforward comparison. More information on the PESs, QCT, and QD calculations for the two systems can be found in earlier publications[8, 37, 41] and in the SI.

It is worthwhile to first examine the internal energy of the gaseous molecule as a function of gas temperature ($T_g$). It should be emphasized that the conventional PIMD theory treats all atoms distinguishable and therefore neglects nuclear spin statistics of ortho/para $H_2$ and $D_2O$. To compare with PIMD results, we calculate quantum internal energies (QIEs) with and without incorporating the intrinsic ratio of the ortho/para states (3:1 for $H_2$ or 2:1 for $D_2O$). As shown in Fig. S6, QIEs of $H_2$ and $D_2O$ obtained in both ways differ only very slightly at low temperature and become indistinguishable at room temperature and above. In Fig. 1, the PIMD internal energies computed via the centroid virial theorem[42] in both molecules are found to gradually increase with the number of beads and converge to QIEs (without nuclear spin statistics) with $P$=40 for $H_2$ and $P$=30 for $D_2O$ at $T_g$>200 K. The convergence becomes increasingly more difficult as $T_g$ decreases (see more clearly in Fig. S5). Indeed, at very low temperature, special symmetry adapted sampling should be applied to distinguish the nuclear spin species along with thousands of beads, as done by Roy and coworkers in path integral Monte Carlo (PIMC) simulations for obtaining superfluid density and related properties at a few Kelvin[43].

In this work, we choose to sample both molecules at a fixed gas temperature ($T_g$=300 K) on a rigid surface for simplicity. So only the ZPE of the molecule is incorporated in the RPMD simulations. Manolopoulos *et al.* found that the rate constants for $F+H_2$ and $H+H_2$ reactions calculated with and without accounting for nuclear spin statistics differ only by 1% or even less at this temperature[27]. Our QD results are almost completely unaffected by nuclear spin statistics. More convergence tests and discussion on nuclear spin statistics can be found in the SI. The reader should note that this gas temperature is not equivalent to the nozzle temperature ($T_n$) used in molecular beam experiments. In practice, one should keep in mind that the incidence energy is often related to $T_n$ in supersonic molecular beams and is subject to a velocity distribution of the carrier gas. This can be modeled by assigning a velocity spread to the initial configurations depending on $T_n$. In addition, while the vibrational temperature is typically close to the nozzle temperature of the beam, the rotational temperature depends on efficiency of rotational cooling. For the $H_2$ beam[44], for example, the rotational temperature is often close to nozzle temperature, or approximately, $T_{rot}\approx0.8\ T_n$. Previous studies have found very similar results by using $T_{rot}\approx T_n$ and $T_{rot}\approx0.8\ T_n$ (*e.g.*, see Ref. 40). In such a case, the current sampling procedure should be appropriate. On the other hand, the rotational temperature is much lower for the $D_2O$ or $CH_4$ beam[3], where only the lowest a few rotational states are populated. This is a challenging case for RPMD simulations, in which the separation the vibrational and rotational sampling may be necessary. Further development is required in this aspect.

Fig. 2 compares $S_0$ values for $H_2$ on Cu(111) as a function of normal incidence energy ($E_i$), obtained by QCT, RPMD and QD calculations. Since our initial state wavepacket method yields the state-specific reaction probability only, the QD results at $T_g$=300 K are obtained by the Boltzmann average of $S_0$ values of low-lying rovibrational states until converged. We have also calculated QCT sticking probabilities at $T_g$=300 K by the sampling of Boltzmann distributions of internal states. The resulting thermally averaged $S_0$ curves at low energies are slightly higher than the ground state reactivity. QCT results agree with QD ones quite well when $E_i$ is above the ZPE corrected minimum barrier height ($E_{bc}$=0.65 eV), consistent with previous findings using a chemically accurate PES by Kroes *et al.*[11, 40]. This implies that the ZPE is well conserved in this highly activated system when $E_i > E_{bc}$ and QCT is a good choice here. However, as $E_i$ decreases to 0.53 eV or lower, the QCT $S_0$ curve drops drastically and becomes substantially lower than the QD counterpart. At $E_i \approx 0.50$ eV, for example, the ground state $S_0$ for QD is ~70 times larger than that for QCT. This failure of QCT is due apparently to the absence of quantum tunneling so that the dissociation channel is suddenly closed when the molecular energy is lower than a threshold. The underestimation of $S_0$ by QCT at $T_g$=300 K is still significant at low $E_i$ and, though less severe. The reason is that the vibrationally adiabatic barrier height becomes lower for internally excited states, making the tunneling effect at the same incidence energy less important, for which QCT does a better job. Note that this tunneling region may shift to lower energy on the chemically accurate PES[40] given its lower barrier, but the phenomenon is supposed to be similar. In contrast, RPMD correctly captures the significant tunneling contribution at very low energies, yielding $S_0$ values much closer to QD ones. However, the agreement between RPMD and QD results is less satisfactory at $E_i$ =0.6~1.0 eV. Similarly, in a recent NE-RPMD study for H scattering on graphene[31], Jiang *et al.* also found that the RPMD calculated sticking probabilities are generally lower than classical ones (note that there is no internal state and ZPE for hydrogen atom), at high incidence energies. As discussed in more detail in the SI, we find that some replicas of the $H_2$ molecule in NE-RPMD simulations may access too close to the surface and feel strong repulsive force to prevent the molecular dissociation. There is no such problem in classical trajectories so that the QCT predicts higher $S_0$ than RPMD does in this energy range. This effect should be less important in the case of $D_2O$ because the heavy oxygen atom (and its replicas) would not approach closely to the surface.

It is important to note that much higher nozzle temperatures are needed to achieve a collision energy of 0.5 eV for $H_2$ ($\approx$2100 K rather than 300 K)[44], where the excited states would indeed largely contribute to the overall reactivity. These excited state contributions would be mainly over-barrier reactivity, making the tunneling effect less prominent. As a result, our conclusions derived for a gas temperature of 300 K might not apply to sticking probabilities in such experiments. However, when the activation barrier is much higher, *e.g.* $H_2$ + Au(111), or the incident molecule is heavier, *e.g.* HCl + Au(111), the tunneling would dominate even at relatively high incidence energies. For instance, in the experiment of Wodtke *et al.* for HCl +Au(111)[45], the sticking probability is lower than $10^{-5}$ at a nozzle temperature of 295 K and an incidence energy of ~1.0 eV, where we expect the tunneling effect would be important and QCT would underestimate the reactivity. This is indeed a possible reason responsible for the large discrepancy between the QCT and QD results in this energy range[46]. In addition, some associative desorption experiments also provide the state-selected sticking probabilities derived from detailed balance, at very low energies below the ZPE-corrected barrier, *e.g.* see $H_2+Cu(111)$[47] and $H_2+Au(111)$[48]. These experiments are typically done at very high surface temperatures, so one may consider using the QCT method which easily incorporates the surface motion. Our results indeed suggest that QCT calculations for this purpose may be used with caution.

Next, we compare RPMD, QCT and QD results for $D_2O+Ni(111)$ in Fig. 3. It is clear that QCT overestimates the sticking probability considerably in the entire energy range, and the overestimation can be more than one order of magnitude near the ZPE corrected barrier ($E_{bc}$=0.58 eV). Analogous behavior has been observed on methane dissociation by Mastromatteo and Jackson[15], when comparing their reaction path Hamiltonian based quasi-classical and quantum dynamical models. In these polyatomic reactions, the reactant ZPE is very large, *e.g.* 0.41 eV for $D_2O$ and ~1.2 eV for $CH_4$[15]. Classical mechanics allows this energy to flow unphysically into the other molecular DOFs, even at the incidence energy higher than $E_{bc}$. Since the reaction probability is relatively low for $D_2O$ on Ni(111) in the energy range of interest, in particular, a small amount of ZPE leaking into the reaction coordinate would lead to a significant increase of reactivity. We note that ZPE conversion is not a severe problem in quasi-classical AIMD simulations for $CHD_3$ dissociation on Ni and Pt surfaces above $E_{bc}$, especially for $CHD_3(v_1=1)$ whose vibrational energy is primarily localized in the C-H bond, alleviating the artificial intramolecular vibrational redistribution[13-14]. Impressively, RPMD does a much better job than QCT in the entire energy range, due to its intrinsic ability to avoid the spurious ZPE leakage. Similarly, Habershon and Manolopoulos found that the linearized semiclassical initial value representation method results in much larger diffusion coefficient of liquid water than the RPMD method, because of the ZPE leakage in the former[30]. We note that Truhlar and coworkers recently proposed an extended Hamiltonian molecular dynamics (EHMD) method that maintains ZPE in the Henon–Heiles model Hamiltonian very well[22]. By connecting two images of a trajectory by one or more springs, this EHMD method can be regarded as a simpler version of RPMD.

Nevertheless, the RPMD $S_0$ values at $T_g$=300 K become increasingly higher QD ones for $D_2O$ in its rovibrationally ground state as $E_i$ decreases. This overestimation may result from the inconsistent comparison between the thermal RPMD results at 300 K and the ground state QD ones. Unfortunately, in this case, there are too many internal states of $D_2O$ having considerable populations that may contribute to the Boltzmann average of $S_0$, making the exact thermally-averaging QD calculations very demanding. To estimate the thermal contribution, we perform additional QCT calculations based on the Boltzmann sampling of internal states at $T_g$=300 K. It is found that the population of rovibrationally excited states leads to a much greater reactivity and the enhancement is larger at a lower incidence energies, confirming the crucial effects of internal excitation. Such effects on the reactivity are often quantified by the horizontal translational energy shift ($\Delta E_i$) between the ground state and the excited state $S_0$ curves[1]. The ratio of the corresponding internal excitation energy ($E_{vj}$) and this shift, *i.e.* $E_{vj}/\Delta E_i$, determines the so-called rovibrational efficacy ($\eta$) representing the effectiveness of the internal energy in promoting the reaction relative to the translational energy[1]. It has been found that the fundamental vibrational excitations of $D_2O$ have relatively high efficacies ($\eta \approx 1.1$~$1.7$ for static surface)[41], while the higher energy overtones and rotational excitations would have lower efficacies[3, 49]. For simplicity, we assume a unified efficacy $\eta = 1.0$ on average and obtain the shifted $S_0$ curves for all populated states, then thermally average them to yield an estimated QD $S_0$ curve at $T_g$=300 K. This averaged QD $S_0$ curve is found to be insensitive to the $\eta$ value in a reasonable range (see Fig. S9b). Alternatively, we assume that QCT and QD methods would have similar thermal contributions due to internal excitation, which is justified by the comparable vibrational efficacies previously predicted by QCT[8] and QD[41] calculations. The relative difference between QCT ($T_g$=300 K) and QCT (ground state) results can be thus added onto the QD ground state $S_0$ curve to approximate the QD $S_0$ curve at $T_g$=300 K. Interestingly, these two estimated thermal QD results are in line with each other, which are both in excellent agreement between the RPMD ones. It should be noted that these two ways to account for thermal effects due to internal state excitations are approximate, but they are not unreasonable. A unified efficacy for all thermally populated states was used before by Jackson and Nave[50]. Compared to the average over explicitly calculated sticking probabilities of the lowest ten initial states only, this approximate treatment was found to yield better agreement with experimental data at various nozzle temperatures. We feel that the current level of comparison is sufficient for our purpose. More accurate individual initial state reaction probabilities can be calculated in future work.

To summarize, we reveal here two striking manifestations of quantum effects in two benchmark surface reactions, which can not be well described by the conventional QCT method. In light of the careful comparison of QCT, RPMD, and QD calculated sticking probabilities, we demonstrate that the NE-RPMD approach can be alternatively used in such cases to capture the salient quantum nature of molecular dissociation on metal surfaces. For $H_2$+Cu(111), NE-RPMD successfully describes the reactivity at very low incidence energies dominated by quantum tunneling, which is largely underestimated by QCT. For $D_2O$ dissociation on Ni(111), QCT suffers from severe unphysical ZPE leakage to the reaction coordinate, resulting in more than one order of magnitude overestimation of dissociation probability even at translational energies above the ZPE corrected barrier. Taking the internal excitation at a given temperature into account, albeit approximately, RPMD results agree rather well with QD ones. Although the NE-RPMD method is validated here within BOSS approximation, it is scalable to include lattice motion given its trajectory-based nature[29], when combined with recently developed molecule-surface PESs explicitly involving the surface DOFs[51]. Incorporating the ZPE of the surface will be another advantage of RPMD superior to QCT. The success of the NE-RPMD method in calculating reaction probability is very encouraging, given its intrinsic ability of dealing with tunneling and quantum ZPE in polyatomic surface reactions. However, more work needs to be done for a more realistic sampling of molecular beam conditions with efficient rotational cooling. NE-RPMD also underestimates the reactivity for $H_2$ on Cu(111) at high incidence energies, due presumably to the fact that some replicas of hydrogen molecule may access the surface too closely. We hope this work will inspire further development of this promising approach for modelling quantum effects in reactive scattering problems.

**Acknowledgements:** Q. L., L. Z., and B. J. were supported by National Key R&D Program of China (2017YFA0303500), National Natural Science Foundation of China (21573203, 91645202, 21722306, 21503130, and 11674212), and Anhui Initiative in Quantum

Information Technologies (AHY090200). Y. L. was supported by the Shanghai Key Laboratory of High Temperature Superconductors (No. 14DZ2260700). We appreciate the Supercomputing Center of USTC for high-performance computing services. We thank Profs. Thomas Miller III and Hua Guo for some helpful discussion.

## References

1.	Juurlink, L. B. F.; Killelea, D. R.; Utz, A. L. State-resolve probes of methane dissociation dynamics. *Prog. Surf. Sci.* **2009,** *84*, 69-134.

2.	Chadwick, H.; Beck, R. D. Quantum state–resolved studies of chemisorption reactions. *Annu. Rev. Phys. Chem.* **2017,** *68* (1), 39-61.

3.	Hundt, P. M.; Jiang, B.; van Reijzen, M.; Guo, H.; Beck, R. D. Vibrationally promoted dissociation of water on Ni(111). *Science* **2014,** *344*, 504-507.

4.	Kroes, G.-J.; Diaz, C. Quantum and classical dynamics of reactive scattering of $H_2$ from metal surfaces. *Chem. Soc. Rev.* **2016,** *45*, 3658-3700.

5.	Jiang, B.; Yang, M.; Xie, D.; Guo, H. Quantum dynamics of polyatomic dissociative chemisorption on transition metal surfaces: Mode specificity and bond selectivity. *Chem. Soc. Rev.* **2016,** *45*, 3621-3640.

6.	Shen, X.; Zhang, D. H. Recent advances in quantum dynamics studies of gas-surface reactions. *Adv. Chem. Phys.* **2018,** *163*, 7.

7.	Gross, A.; Wilke, S.; Scheffler, M. Six-dimensionalquantum dynamics of adsorption and desorption of $H_2$ at Pd(100): Steering and steric effects. *Phys. Rev. Lett.* **1995,** *75*, 2718-2721.

8.	Jiang, B.; Guo, H. Dynamics of water dissociative chemisorption on Ni(111): Effects of impact sites and incident angles. *Phys. Rev. Lett.* **2015,** *114*, 166101.

9.	Shen, X.; Chen, J.; Zhang, Z.; Shao, K.; Zhang, D. H. Methane dissociation on Ni(111): A fifteen-dimensional potential energy surface using neural network method. *J. Chem. Phys.* **2015,** *143* (14), 144701.

10.	Zhang, Z.; Liu, T.; Fu, B.; Yang, X.; Zhang, D. H. First-principles quantum dynamical theory for the dissociative chemisorption of $H_2O$ on rigid Cu(111). *Nat. Comm.* **2016,** *7*, 11953.

11.	Diaz, C.; Olsen, R. A.; Busnengo, H. F.; Kroes, G.-J. Dynamics on six-dimensional potential energy surfaces for $H_2$/Cu(111): Corrugation reducing procedure versus modified Shepard interpolation method and PW91 versus RPBE. *J. Phys. Chem. C* **2010,** *114* (25), 11192-11201.

12.	Groß, A. Ab Initio Molecular Dynamics Study of Hot Atom Dynamics after Dissociative Adsorption of $H_2$ on Pd(100). *Phys. Rev. Lett.* **2009,** *103* (24), 246101.

13.	Nattino, F.; Ueta, H.; Chadwick, H.; van Reijzen, M. E.; Beck, R. D.; Jackson, B.; van Hemert, M. C.; Kroes, G.-J. Ab initio molecular dynamics calculations versus quantum-state-resolved experiments on $CHD_3$ + Pt(111): New insights into a prototypical gas–surface reaction. *J. Phys. Chem. Lett.* **2014,** *5* (8), 1294-1299.

14.     Nattino, F.; Migliorini, D.; Kroes, G.-J.; Dombrowski, E.; High, E. A.; Killelea, D. R.; Utz, A. L. Chemically accurate simulation of a polyatomic molecule-metal surface reaction. *J. Phys. Chem. Lett.* **2016,** *7* (13), 2402-2406.

15.     Mastromatteo, M.; Jackson, B. The dissociative chemisorption of methane on Ni(100) and Ni(111): Classical and quantum studies based on the reaction path Hamiltonian. *J. Chem. Phys.* **2013,** *139* (19), 194701.

16.     Zhou, X.; Jiang, B. Mode-specific and bond-selective dissociative chemisorption of CHD3 and CH2D2 on Ni(111) revisited using a new potential energy surface. *Science China Chemistry* **2018,** *61* (9), 1134-1142.

17.     Gross, A.; Scheffler, M. Ab initio quantum and molecular dynamics of the dissociative adsorption of hydrogen on Pd(100). *Phys. Rev. B* **1998,** *57*, 2493.

18.     Busnengo, H. F.; Crespos, C.; Dong, W.; Rayez, J. C.; Salin, A. Classical dynamics of dissociative adsorption for a nonactivated system: The role of zero point energy. *J. Chem. Phys.* **2002,** *116* (20), 9005-9013.

19.     Miller, W. H.; Hase, W. L.; Darling, C. L. A simple model for correcting the zero point energy problem in classical trajectory simulations of polyatomic molecules. *J. Chem. Phys.* **1989,** *91*, 2863-2868.

20.     Bowman, J. M.; Gazdy, B.; Sun, Q. A method to constrain vibrational energy in quasiclassical trajectory calculations. *J. Chem. Phys.* **1989,** *91*, 2859-2862.

21.     Paul, A. K.; Hase, W. L. Zero-point energy constraint for unimolecular dissociation reactions. Giving trajectories multiple chances to dissociate correctly. *J. Phys. Chem. A* **2016,** *120* (3), 372-378.

22.     Shu, Y.; Dong, S. S.; Parker, K. A.; Bao, J. L.; Zhang, L.; Truhlar, D. G. Extended Hamiltonian molecular dynamics: semiclassical trajectories with improved maintenance of zero point energy. *Phys. Chem. Chem. Phys.* **2018,** *20* (48), 30209-30218.

23.     Varandas, A. J. C. A novel non-active model to account for the leak of zero-point energy in trajectory calculations. Application to H + $O_2$ reaction near threshold. *Chem. Phys. Lett.* **1994,** *225*, 18-27.

24.     Habershon, S.; Manolopoulos, D. E.; Markland, T. E.; Miller III, T. F. Ring-polymer molecular dynamics: Quantum effects in chemical dynamics from classical trajectories in an extended phase space. *Annu. Rev. Phys. Chem.* **2013,** *64*, 387-413.

25.     Craig, I. R.; Manolopoulos, D. E. Quantum statistics and classical mechanics: Real time correlation function from ring polymer molecular dynamics. *J. Chem. Phys.* **2004,** *121*, 3368-3373.

26.     Craig, I. R.; Manolopoulos, D. E. Chemical reaction rates from ring polymer molecular dynamics. *J. Chem. Phys.* **2005,** *122*, 084106.

27.     Collepardo-Guevara, R.; Suleimanov, Y. V.; Manolopoulos, D. E. Bimolecular reaction rates from ring polymer molecular dynamics. *J. Chem. Phys.* **2009,** *130*, 174713.

28.     Suleimanov, Y. V.; Aoiz, F. J.; Guo, H. Chemical reaction rate coefficients from ring polymer molecular dynamics: Theory and practical applications. *J. Phys. Chem. A* **2016,** *120* (43), 8488-8502.

29.     Suleimanov, Y. V. Surface diffusion of hydrogen on Ni(100) from ring polymer molecular dynamics. *J. Phys. Chem. C* **2012,** *116*, 11141-11153.

30.     Welsch, R.; Song, K.; Shi, Q.; Althorpe, S. C.; Miller, T. F. Non-equilibrium dynamics from RPMD and CMD. *J. Chem. Phys.* **2016,** *145* (20), 204118.

31.     Jiang, H.; Kammler, M.; Ding, F.; Dorenkamp, Y.; Manby, F. R.; Wodtke, A. M.; Miller, T. F.; Kandratsenka, A.; Bünermann, O. Imaging covalent bond formation by H atom scattering from graphene. *Science* **2019,** *364* (6438), 379.

32.     Suleimanov, Y. V.; Aguado, A.; Gómez-Carrasco, S.; Roncero, O. A ring polymer molecular dynamics approach to study the transition between statistical and direct mechanisms in the $H_2 + H_3^+ \rightarrow H_3^+ + H_2$ reaction. *J. Phys. Chem. Lett.* **2018,** *9*, 2133-2137.

33.     Chandler, D.; Wolynes, P. G. Exploiting the isomorphism between quantum theory and classical statistical mechanics of polyatomic fluids. *J. Chem. Phys.* **1981,** *74*, 4078-4095.

34.     Habershon, S.; Manolopoulos, D. E. Zero point energy leakage in condensed phase dynamics: an assessment of quantum simulation methods for liquid water. *J. Chem. Phys.* **2009,** *131* (24), 244518.

35.     Richardson, J. O.; Althorpe, S. C. Ring-polymer molecular dynamics rate-theory in the deep-tunneling regime: Connection with semi-classical instanton theory. *J. Chem. Phys.* **2009,** *131*, 214106.

36.     Pérez, A.; Tuckerman, M. E.; Müser, M. H. A comparative study of the centroid and ring-polymer molecular dynamics methods for approximating quantum time correlation functions from path integrals. *J. Chem. Phys.* **2009,** *130* (18), 184105.

37.     Dai, J.; Zhang, J. Z. H. Quantum adsorption dynamics of a diatomic molecule on surface: Four-dimensional fixed-site model for $H_2$ on Cu(111). *J. Chem. Phys.* **1995,** *102* (15), 6280-6289.

38.     Liu, T.; Fu, B.; Zhang, D. H. Validity of the site-averaging approximation for modeling the dissociative chemisorption of $H_2$ on Cu(111) surface: A quantum dynamics study on two potential energy surfaces. *J. Chem. Phys.* **2014,** *141*, 194302.

39.     Dai, J.; Light, J. C. The steric effect in a full dimensional quantum dynamics simulation for the dissociative adsorption of $H_2$ on Cu(111). *J. Chem. Phys.* **1998,** *108*, 7816-7820.

40.     Díaz, C.; Pijper, E.; Olsen, R. A.; Busnengo, H. F.; Auerbach, D. J.; Kroes, G.-J. Chemically accurate simulation of a prototypical surface reaction: $H_2$ dissociation on Cu(111). *Science* **2009,** *326*, 832-834.

41.     Jiang, B.; Song, H.; Yang, M.; Guo, H. Quantum dynamics of water dissociative chemisorption on rigid Ni(111): An approximate nine-dimensional treatment. *J. Chem. Phys.* **2016,** *144* (16), 164706.

42.     Herman, M. F.; Bruskin, E. J.; Berne, B. J. On path integral Monte Carlo simulations. *J. Chem. Phys.* **1982,** *76* (10), 5150-5155.

43.     Zeng, T.; Blinov, N.; Guillon, G.; Li, H.; Bishop, K. P.; Roy, P.-N. MoRiBS-PIMC: A program to simulate molecular rotors in bosonic solvents using path-integral Monte Carlo. *Comput. Phys. Commun.* **2016,** *204*, 170-188.

44.     Rettner, C. T.; Michelsen, H. A.; Auerbach, D. J. Quantum-state-specific dynamics of the dissociative adsorption and associative desorption of $H_2$ at a Cu(111) surface. *J. Chem. Phys.* **1995,** *102* (11), 4625-4641.

45.     Shirhatti, P. R.; Geweke, J.; Steinsiek, C.; Bartels, C.; Rahinov, I.; Auerbach, D. J.; Wodtke, A. M. Activated dissociation of HCl on Au(111). *J. Phys. Chem. Lett.* **2016,** *7* (7), 1346-1350.

46.     Liu, Q.; Zhou, X.; Zhou, L.; Zhang, Y.; Luo, X.; Guo, H.; Jiang, B. Constructing high-dimensional neural network potential energy surfaces for gas–surface scattering and reactions. *J. Phys. Chem. C* **2018,** *122* (3), 1761-1769.

47.     Kaufmann, S.; Shuai, Q.; Auerbach, D. J.; Schwarzer, D.; Wodtke, A. M. Associative desorption of hydrogen isotopologues from copper surfaces: Characterization of two reaction mechanisms. *J. Chem. Phys.* **2018,** *148* (19), 194703.

48.     Shuai, Q.; Kaufmann, S.; Auerbach, D. J.; Schwarzer, D.; Wodtke, A. M. Evidence for Electron-Hole Pair Excitation in the Associative Desorption of H2 and D2 from Au(111). *J. Phys. Chem. Lett.* **2017,** *8* (7), 1657-1663.

49.     Jiang, B. Rotational and steric effects in water dissociative chemisorption on Ni(111). *Chem. Sci.* **2017,** *8* (9), 6662-6669.

50.     Jackson, B.; Nave, S. The dissociative chemisorption of methane on Ni(111): The effects of molecular vibration and lattice motion. *J. Chem. Phys.* **2013,** *138*, 174705.

51.     Zhang, Y.; Zhou, X.; Jiang, B. Bridging the Gap between Direct Dynamics and Globally Accurate Reactive Potential Energy Surfaces Using Neural Networks. *J. Phys. Chem. Lett.* **2019,** *10*, 1185-1191.

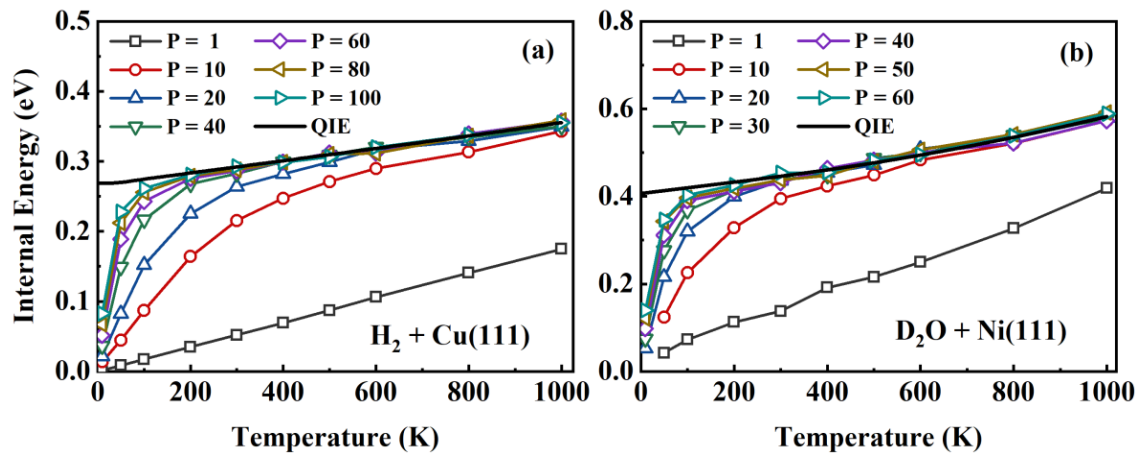

Fig. 1. PIMD internal energies of $H_2$ (a) and $D_2O$ (b) estimated from NVT simulations of $H_2$+Cu(111) and $D_2O$+Ni(111) systems with the increasing number of beads ($P$), as a function of molecular temperature. The exact quantum internal energy (QIE) serves as a benchmark for each system, which is evaluated by the thermally averaged energy of all internal states. Nuclear spin statistics has a invisible effect in this plot. See the SI for more details.

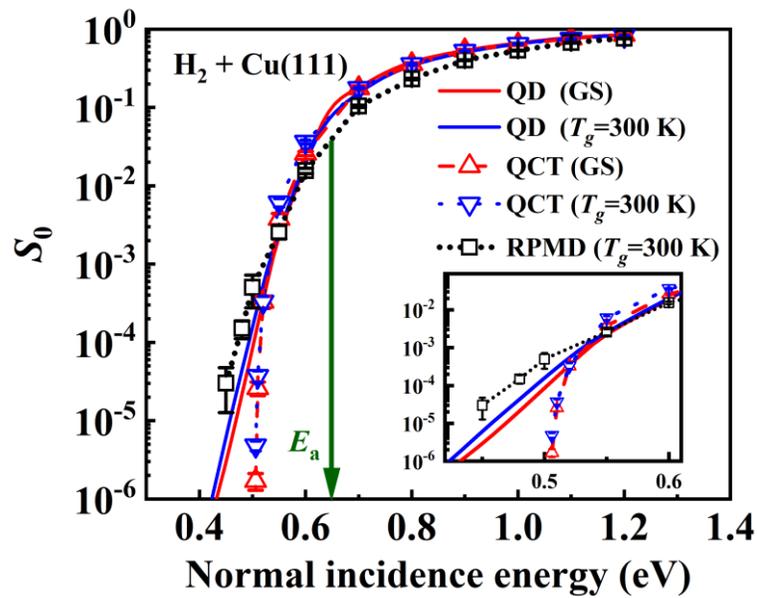

Fig. 2 Comparison of $S_0$ of $H_2$ on Cu(111) obtained from QD for the ground state (GS, red solid line), Boltzmann-averaged QD results at $T_g$=300 K (blue solid line), QCT for ground state (GS, red upper triangle), Boltzmann-averaged QCT results at $T_g$=300 K (blue lower triangle), and RPMD at $T_g$=300 K (black square). The inset zooms in the very low energy regime and the green arrow indicates the ZPE corrected barrier height $E_{bc}$=0.65 eV.

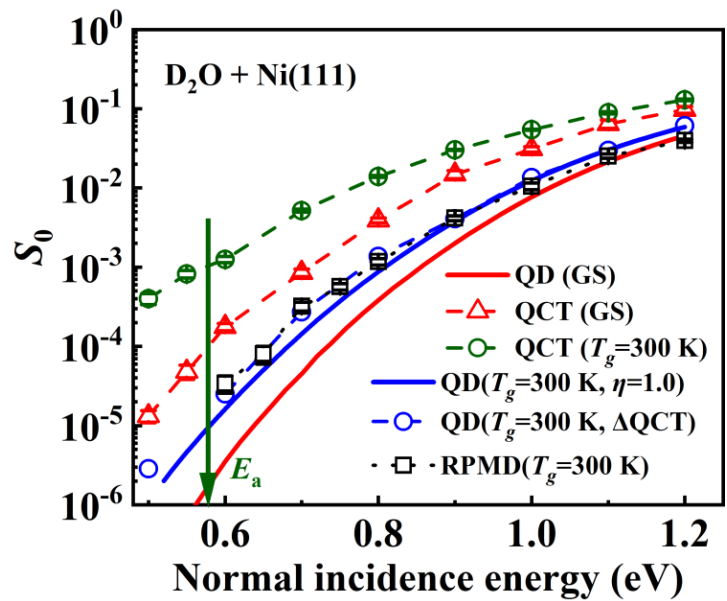

Fig. 3 Comparison of $S_0$ of $D_2O$ on Ni(111) obtained from QD for the ground state (GS, red solid line), QCT for the ground state (GS, red triangle), QCT sampled from Boltzmann distribution at $T_g$=300 K (green circle), QD at $T_g$=300 K estimated by rovibrational efficacy ($\eta$=1.0, blue solid line), QD at $T_g$=300 K estimated by QCT thermal contribution ($\Delta$QCT, blue circle) and RPMD at $T_g$=300 K (black square). The green arrow indicates the ZPE corrected barrier height $E_{bc}$=0.58 eV.

**Supporting Information**

**Ring Polymer Molecular Dynamics in Gas Surface Reactions: Inclusion of Quantum Effects Made Simple**

Qinghua Liu[1], Liang Zhang[1], Yongle Li[2,*], Bin Jiang[1,*]

*[1]Hefei National Laboratory for Physical Science at the Microscale, Department of Chemical Physics, Key Laboratory of Surface and Interface Chemistry and Energy Catalysis of Anhui Higher Education Institutes, University of Science and Technology of China, Hefei, Anhui 230026, China*

*[2] Department of Physics, International Center of Quantum and Molecular Structures and Shanghai Key Laboratory of High Temperature Superconductors, Shanghai University, Shanghai 200444, China.*

*: corresponding authors: yongleli@shu.edu.cn, bjiangch@ustc.edu.cn

# I.  Computational Details

## A. Potential energy surfaces

Two benchmark systems have been tested in this work, namely $H_2$+Cu(111) and $D_2O$+Ni(111). Within the Born-Oppenheimer static surface (BOSS) approximation, the corresponding potential energy surfaces (PESs) are represented by six and nine molecular degrees of freedom (DOFs), respectively. The Jacobi coordinates of the two systems are depicted in Fig. S1. For $H_2$+Cu(111), we use an analytical London-Eyring-Polanyi-Sato (LEPS) PES of Dai and Zhang[1] that was parameterized with limited density functional theory (DFT) data. The DFT calculations were based on the generalized gradient approximation (GGA)[2] (but with the densities calculated from local density approximation) and pseudopotentials with a four layer Cu(111) slab. The classical dissociation barrier height for bridge site is lowest (0.72 eV), followed by the hollow (0.90 eV) and top sites (1.14 eV). Including zero point energies (ZPEs) of the reactant and transition state, the lowest ZPE corrected barrier over the bridge site becomes 0.65 eV. A two-dimensional (2D) cut of the PES is displayed in Fig. S2a, as a function of $Z$ and $r$ with the molecule parallel to the surface at the bridge site dissociating to the two hollow sites.

For $D_2O$+Ni(111), there has been a first-principles nine-dimensional (9D) PES of Jiang and Guo[3] that was accurately fitted to over 25000 DFT data in terms of the permutation invariant polynomial neural network (PIP-NN) approach[4]. The DFT data was generated at the GGA level with the Perdew-Wang (PW91) functional[2] using a 4-layer Ni(100) slab and 3×3 unit cell. The site-specific classical barrier height on this PES ranges from 0.67 to 0.87 eV[3]. The lowest transition state is nearby the top site and the ZPE corrected activation barrier is 0.58 eV[3]. A 2D PES contour plot is shown in Fig. S2b, as a function of $Z$ and $r_2$ with molecular center of mass (COM) fixed at the transition state and other angular DOFs optimized. More details of the two PESs can be found in the original publications[1, 3].

## B. Quasi classical trajectory

The quasi classical trajectory (QCT) method is a standard approach to involve the initial quantization of reactants in molecular dynamics simulations. Our QCT calculations were performed with the VENUS program[5] heavily modified by us for gas-surface scattering[4]. For $H_2$+Cu(111), the $H_2$ molecule is treated as rotating oscillator, whose internal energy is determined by Einstein–Brillouin–Keller (EBK) semiclassical method[6] with given vibrational and rotational quantum numbers $v$ and $j$. Specifically, the EBK semi-classical method gives us the rovibrational energy of $H_2$ ($E_{vj}$) and the inner ($r_-$) and outer ($r_+$) turning points. A random distance $r$ within $[r_-, r_+]$ interval is selected, and its corresponding vibrational momentum is calculated by $p_r = \sqrt{2\mu\left[E_{vj} - E_r(r) - V(r)\right]}$, where $E_r(r) = j^2 / 2\mu r^2$ and $V(r)$ are the rotational and potential energy of $H_2$ at a bond length of $r$, $\mu$ is the reduced mass. Since the probability of a particular $r$ is proportional to the inverse of $p_r$[7], namely $P(r) \propto 1/p_r$, an $(r, p_r)$ pair is accepted only if $P(r)/P^0(r) > R$, where $R$ is a random number, and $P^0(r)$ is calculated by $p_r$ for the most probable $r$ (i.e. the turning point)[8]. The vibrational phase is determined by another random number $R'$, whether $R' > 0.5$ (positive) or $R' < 0.5$ (negative). Indeed, the EBK semiclassical quantization for $H_2$ reproduces the quantum energy levels quite accurately (see Table S1). The initial $(X, Y)$ coordinates of the $H_2$ COM was randomly chosen in a unit cell and $Z$ is fixed at 8 Å above the surface. The initial polar and azimuthal angles $(\theta, \varphi)$ are also randomly sampled. The initial COM translational momentum is along surface normal. A trajectory is deemed as reactive if the H-H distance is larger than 2.5 Å and as scattered if the molecule is 8.1 Å above the surface with a velocity away from the surface.

For $D_2O+Ni(111)$, the molecular COM and orientation are sampled in the same way as for $H_2+Cu(111)$. Standard normal mode sampling is employed to prepare the ground state of $D_2O$ or a collection of internal states based on the Boltzmann probabilities of them at given temperature ($T_g = 300$ K, here). A trajectory is deemed reactive if the longer O-D distance is larger than 2.2 Å and scattering if the molecule is 8.1 Å above the surface with a velocity away from the surface. In both systems, the propagation is performed with a time interval of 0.1 fs using the velocity Verlet algorithm until the trajectory reaches either the dissociation or the scattered channel. The initial sticking probability ($S_0$) can be calculated as $S_0=N_r/N_{tot}$, where $N_r$ represents the number of reactive trajectory, and $N_{tot}$ is the number of total effective trajectory.

## C. Ring polymer molecular dynamics

As discussed in the main text, the ring polymer molecular dynamics (RPMD) theory is established on an isomorphism from the equilibrium quantum mechanical partition function to the classical partition function of a fictitious ring polymer[9]. The corresponding equilibrium RPMD Hamiltonian is extracted from the expression of partition function as[10],

$$H_P(\mathbf{p}, \mathbf{q}) = \sum_{k=1}^{P} \sum_{j=1}^{N} \left[ \frac{(\mathbf{p}_j^k)^2}{2m_j} + \frac{1}{2} m_j \omega_P^2 (\mathbf{q}_j^k - \mathbf{q}_j^{k-1})^2 \right] + \sum_{k=1}^{P} V(\mathbf{q}_1^k, \mathbf{q}_2^k, ..., \mathbf{q}_N^k) \qquad (1)$$

where $N$ and $P$ are the number of atoms in the system and beads in the ring polymer, $\mathbf{p}_j^k$ and $\mathbf{q}_j^k$ are the momentum and position vectors of the $k$th bead of the ring polymer representing the $j$th atom and $\mathbf{q}_j^0 = \mathbf{q}_j^P$, $V$ is the PES depending on the positions of the same numbering of the beads for all atoms, $m_j$ is the $j$th atomic mass that is used for each bead of this atom, and $\omega_P$ is the frequency harmonic potential taking the value of $\omega_P = P / \beta$ ($\beta = 1/kT$). The time evolution of the RPMD Hamiltonian is then fully determined by the classical dynamics so that the coordinate and momentum of each bead are propagated by,

$$\dot{\mathbf{p}}_j^k = \frac{\partial H_P(\mathbf{p}, \mathbf{q})}{\partial \mathbf{q}_j^k}$$

$$\dot{\mathbf{q}}_j^k = \frac{\partial H_P(\mathbf{p}, \mathbf{q})}{\partial \mathbf{p}_j^k} \qquad (2)$$

The corresponding RPMD time correlation function in equilibrium conditions well approximates the quantum counterpart and even becomes exact in several important limits, including short time, high temperature, and harmonic interaction potentials[11].

For describing the gas-surface collision process with a specific incidence energy, we have to use the non-equilibrium RPMD (NE-RPMD) Hamiltonian[12],

$$H'_P(\mathbf{p}, \mathbf{q}) = \sum_{k=1}^{P} \sum_{j=1}^{N} \left[ \frac{(\mathbf{p}_j^k - \Delta\mathbf{p})^2}{2m_j} + \frac{1}{2} m_j \omega_P^2 (\mathbf{q}_j^k - \mathbf{q}_j^{k-1})^2 \right] + \sum_{k=1}^{P} V(\mathbf{q}_1^k, \mathbf{q}_2^k, ..., \mathbf{q}_N^k) \qquad (3)$$

where an initial momentum impulse $\Delta p$ is added that corresponds to the initial translational energy given in molecular beam experiments. Miller and coworkers[12] have shown that several limits of equilibrium RPMD time correlation functions are preserved in non-equilibrium conditions and RPMD is able to achieve similar accuracy for calculating equilibrium and non-equilibrium correlation functions. For our purpose, it is also important that RPMD intrinsically incorporates zero point energy (ZPE) of the system, preserves the quantum Boltzmann distribution at a given temperature[11], and avoids unphysical ZPE leakage by construction[13]. One can estimate the internal energy of the ring polymer system from the centroid virial theorem[14],

$$< E > = \frac{3N}{2\beta} + < \frac{1}{2P} \sum_{k=1}^{P} \sum_{j=1}^{N} (\mathbf{q}_j^k - \bar{\mathbf{q}}_j) \frac{\partial V(\mathbf{q}_1^k, \mathbf{q}_2^k, ..., \mathbf{q}_N^k)}{\partial \mathbf{q}_j^k} > + < \frac{1}{P} \sum_{k=1}^{P} V(\mathbf{q}_1^k, \mathbf{q}_2^k, ..., \mathbf{q}_N^k) > \qquad (4)$$

where $\bar{\mathbf{q}}_j = \sum_{k=1}^{P} \frac{\mathbf{q}_j^k}{P}$ representing the position of centroid.

Our NE-RPMD simulations of gas-surface reactions are implemented in two stages, namely thermalization (NVT ensemble) of the reactants and collision dynamics (NVE ensemble), following the recent work of Suleimanov and coworker[15]. We first run equilibrium path integral molecular dynamics (PIMD) of the gaseous reactant coupled with a thermostat at a given gas temperature ($T_g = 300$ K, here). Here, the bead positions are assembled in the equilibrium geometry of the molecule far above the surface and their momenta are randomly sampled from the Maxwell-Boltzmann distribution at $T_g =$ 300 K. The trajectory is propagated with velocity Verlet algorithm and a time step of 0.1 fs. These beads become naturally scattered once propagated and form ring polymers, because of the inter-bead harmonic potentials. The momentum of each bead is resampled every 10 fs due to the Andersen thermostat[16]. Note that the COM translational energy of the reactant is removed, which is equivalent to imposing a constraint on the reaction coordinate as in the calculation of rate constants[17]. The system accommodates to the given temperature typically within 1 ps, after which the trajectory snapshots are stored every 3 fs (30 steps). This equilibration procedure is an essential stage to ensure ergodicity in PIMD[18]. As a result, these snapshots include bead positions and momenta that can be used as initial conditions for the subsequent collision dynamics simulation. In addition, the molecular COM is randomly placed in the unit cell at $Z = 8$ Å above the surface. An additional momentum pointing to the surface along surface normal is added to each bead for a given normal incidence energy (or the translational energy) $E_n$, namely $\Delta \mathbf{p}_j^k = -m_j \sqrt{2E_i / M}$. These trajectories are then propagated without a thermostat and each atom is assumed to be located at the centroid of the corresponding ring polymer so as to decide whether the trajectory is reactive or not. The criteria for reactive and scattered trajectories are exactly the same as those given above for QCT calculations.

## D. Quantum dynamics

Our quantum dynamics (QD) calculations have been done with the time-dependent wavepacket method, which has been detailed in Refs. [19] and [20]. Briefly, within the BOSS approximation, the quantum wave packet is represented by six and nine Jacobi coordinates in the $H_2$+Cu(111) and $D_2O$+Ni(111) systems, as illustrated in Fig. S1. The uniform quantum Hamiltonian of a molecule-surface system can be written as,

$$\hat{H} = \hat{K}_{trans} + \hat{K}_{vib} + \hat{K}_{rot} + \hat{V}(\mathbf{q}), \tag{5}$$

where $\hat{K}_{trans}$ is the three-dimensional kinetic energy operator (KEO) for the center-of-mass translation ($\hbar=1$ hereafter),

$$
\begin{aligned}
\hat{K}_{trans} &= -\frac{1}{2M}\left(\frac{\partial^2}{\partial Z^2} + \frac{\partial^2}{\partial X^2} + \frac{\partial^2}{\partial Y^2}\right) \\
&= -\frac{1}{2M}\left(\frac{\partial^2}{\partial Z^2} + \frac{1}{\sin^2\gamma}\frac{\partial^2}{\partial u^2} + \frac{1}{\sin^2\gamma}\frac{\partial^2}{\partial v^2} - \frac{2\cos\gamma}{\sin^2\gamma}\frac{\partial^2}{\partial u\partial v}\right).
\end{aligned}
\tag{6}
$$

Here, $Z$ and ($X$, $Y$) correspond to perpendicular and parallel translation of the molecule, $\gamma$ is the skew angle between two lattice vectors ($u$, $v$) and $M$ is the molecular mass. On the other hand, $\hat{K}_{vib}$ and $\hat{K}_{rot}$ represent the vibrational and rotational KEOs of the molecule, respectively. The vibrational KEO depends on both the intramolecular radial and angular coordinates, while the rotational KEO depends on angular variables that describe the rotation and orientation of the molecule, as illustrated in Fig. S1. $\hat{V}(\mathbf{q})$ is the molecule-surface interaction PES depending on all coordinates. Specifically, for the $H_2$+Cu(111) system, $\hat{K}_{vib} + \hat{K}_{rot}$ can be expressed as,

$$\hat{K}_{vib} + \hat{K}_{rot} = -\frac{1}{2\mu}\frac{\partial^2}{\partial r^2} + \frac{\hat{j}^2}{2\mu r^2}, \tag{7}$$

where $\mu$ is the reduced mass of $H_2$, $\hat{j}$ is the angular momentum of $H_2$, $r$ is the H-H

bond length. For the $D_2O$+Ni(111) system, , $\hat{K}_{vib} + \hat{K}_{rot}$ can be expressed as,

$$\hat{K}_{vib} + \hat{K}_{rot} = -\frac{1}{2\mu_1}\frac{\partial^2}{\partial r_1^2} - \frac{1}{2\mu_2}\frac{\partial^2}{\partial r_2^2} + \frac{\hat{j}^2}{2\mu_1 r_1^2} + \frac{(\hat{J}-\hat{j})^2}{2\mu_2 r_2^2}, \qquad (8)$$

where $r_1$ is the bond length of the non-dissociative OD bond, $r_2$ the distance between COM of OD and D, the corresponding reduced masses are $\mu_1 = m_D m_O /(m_D + m_O)$, $\mu_2 = m_{OD} m_D /(m_{OD} + m_D)$, $\hat{j}$ and $\hat{J}$ are the OD and $D_2O$ angular momentum operators, respectively.

The wave function is expanded by the radial and rotational basis functions. Specifically, sine basis functions are used for the translational coordinate $Z$, which is divided into the interaction and asymptotic regions, in order to take advantage of an $L$ grid saving scheme. Periodic Fourier functions are used for for $u$ and $v$. The vibrational basis along $r_i$ ($r$ for $H_2$+Cu system and $i = 1$ or 2 for $D_2O$+Ni system) consists of the reference vibrational eigenfunctions $\phi_{n_i}(r_i)$ for the one-dimensional reference Hamiltonian for the isolated molecule far from the surface, which satisfy the equation,

$$\left[ -\frac{1}{2\mu_i}\frac{\partial^2}{\partial r_i^2} + V_{r_i}(Z_\infty, r_i, \dots) \right] \phi_{n_i}(r_i) = \varepsilon_n \phi_{n_i}(r_i). \qquad (9)$$

For the $H_2$+Cu(111) system, a non-direct product FBR consisting of spherical harmonics $Y_j^{m_j}(\theta, \varphi)$ is employed to represent the angular wave function. While for the $D_2O$+Ni(111) system, The overall rotational basis is described by $Y_{jl}^{JM}(\theta_1, \theta_2, \varphi, \phi)$, which is defined as,

$$Y_{jl}^{JM}(\theta_1, \theta_2, \varphi, \phi) = \sum_K D_{MK}^{J*}(\phi, \theta_2, \varphi)\sqrt{\frac{2l+1}{4\pi}} < jKl0 \,|\, JK > y_{jK}(\theta_1, 0), \qquad (10)$$

where $D^J_{MK}(\phi, \theta_2, \varphi)$ is the Wigner rotation matrix which can be expressed as $\sqrt{\dfrac{2J+1}{8\pi^2}} e^{-iM\phi} d^J_{MK}(\theta_2) e^{-iK\varphi}$, $y_{jK}(\theta_1, 0)$ is the spherical harmonics. $K$ and $M$ are the projections of $J$ on $r_2$ and the surface normal, respectively.

The initial normal-incident wave packet located in the reactant asymptote ($Z = Z_i$) as a product of a Gaussian wave packet in $Z$ and internal state wavefunction of the reactant $\left| \psi_{\text{int}} \right\rangle$,

$$\left| \Psi_i(t=0) \right\rangle = N e^{-(Z-Z_i)^2/2\delta^2} e^{-ik_iZ} \left| \psi_{\text{int}} \right\rangle, \tag{11}$$

where the wave vector $k_i$ is determined by the incidence energy $E_i$: $k_i = \sqrt{2ME_i}$, $Z_i$ and $\delta$ are the initial central position and width of the Gaussian wave packet. The wave function was propagated using the split-operator method and absorbing potential was imposed at the edges of the grid to avoid spurious reflections,

$$D(\zeta) = \exp\left[ -C_\zeta \left( \frac{\zeta - \zeta_a}{\zeta_{\max} - \zeta_a} \right)^2 \right], \quad \zeta = Z \ or \ r_i. \tag{12}$$

The initial state-selected reaction probability ($P_0$, or $S_0$ for molecular initial sticking probability) is obtained by evaluating the energy dependent reactive flux at the dividing surface ($r = r_f$, $r$ is the dissociation bond),

$$P_0(E) = \frac{1}{\mu} \text{Im} \left\langle \psi^+_{iE} \left| \delta(r - r_f) \frac{\partial}{\partial r} \right| \psi^+_{iE} \right\rangle_{r=r_f}, \tag{13}$$

where $\left| \psi^+_{iE} \right\rangle$ is the time-independent wave function which is related to time-dependent wave packet by a half Fourier transformation.

It should be noted that the nine-dimensional QD calculations for the D$_2$O+Ni(111)

system are still extremely expensive. Instead, we take advantage of the site-averaging model proposed by Zhang and coworkers, which has been extensively tested for HCl+Au(111)[21], $H_2$ on Cu(111) and Ag(111)[22-23], and $H_2O$+Cu(111)[24] and $H_2O$+Ni(111)[25]. A simpler but relevant sudden approximation for averaging the site specific reactivity has already been applied by Jackson and coworkers in their reaction path Hamiltonian calculations even earlier[26]. In these activated systems, it has been validated that the exact full-dimensional dissociation probabilities can be well represented by the weighted sum of reduced-dimensional fixed-site ones with a sufficiently larger number of sites. The two sets of results are almost identical, implying negligible errors of this site-averaging approximation. The validity of the site-averaged model is on the basis of the fact that the dynamical steering is minor in these activated systems[23]. In our $D_2O$+Ni(111) PES, two hollow sites, $i.e.$ the fcc and hcp sites, are indistinguishable due to the approximate $C_{6v}$ symmetry enforced. This approximation, which is justified because of the negligibly small energy differences between the two sites[25], allows us to focus on an irreducible triangular region of the unit cell, shown in Fig. S1c. We compute the fixed-site seven-dimensional QD reaction probabilities up to 15 sites displayed in Fig. S1c to converge the nine-dimensional $S_0$ for $D_2O$ on Ni(111).

To achieve numerical convergence, for $H_2$+Cu(111), the two dimensional unit cell ($u$, $v$) is covered by a 14 × 14 evenly spaced Fourier grid. We use 127 sine basis functions ranging from 1.0 to 16.0 bohr for $Z$ with 35 basis functions in the interaction region. 6 vibrational basis functions for $r$ are used in the asymptotic region, while 30 in interaction region ranging from 0.5 to 6.0 bohr. The rotational basis is determined by $j_{max} = 30$, $m_{j_{max}} = 20$. The imaginary absorbing potentials are placed in the range of Z between 12.0 and 16.0 bohr and $r$ between 4.0 and 6.0 bohr, respectively. The dissociation flux is calculated on the dividing surface of $r = 3.5$ bohr. The time step is 10 a.u. and we propagate the wave packets for 15000 a.u. of time to converge the dissociation probabilities. For $D_2O+Ni(111)$, a total of 300 sine basis functions ranging from 2.0 to 17.0 bohr are used for Z with 130 basis functions in the interaction region, and 6 vibrational basis functions for both $r_1$ and $r_2$ were used in the asymptotic region, while 30 for $r_2$ in interaction region ranging from 1.0 to 5.5 bohr allowing for dissociation. The rotational basis set is defined by $J_{max} = 37$, $l_{max} = 35$, $K_{max} = 27$ and $M_{max} = 12$. The flux is analyzed at $r_{2_f} = 3.6$ bohr and the damping function that absorbs the wave packet at the edges of the grid starts from $Z = 15.0$ bohr and $r_2 = 3.7$ bohr. The time step for the propagation is 10 a.u. and the wave packets is propagated up to 22000 a.u. in order to converge the reaction probability. We note that the final reaction probabilities obtained in the wave packet calculations are multiplied by a factor of 2 to account for the two equivalent O-D bonds.

The initial state selected wavepacket method only yields the reaction probability for a specific initial state. To get the thermally averaged reaction probability, one needs to consider the Boltzmann distribution of internal states at a given temperature $T_g$. The thermally averaged reaction probability can be written as[27],

$$P(E_i;T_g) = \sum_{v,j} F_B\left(v,j;T_g\right) P_0\left(v,j;E_i\right).$$  (14)

Here, $P_0\left(v,j;E_i\right)$ is the initial state selected reaction probability and $F_B\left(v,j;T_g\right)$ is the Boltzmann weight of each rovibrational state, which is given by,

$$F_B\left(v, j; T_g\right) = w\left(j\right)(2j+1)\exp\left(-\beta E_{vj}\right)\Big/N \,, \tag{15}$$

where the normalization factor $N$ is the sum of all $F_B\left(v, j; T_n\right)$ terms, $E_{vj}$ is the rovibrational energy relative to the ground state. In particular, $w\left(j\right)$ is a degeneracy factor due to the nuclear spin statistics of $H_2$ and $D_2O$, which is correlated with the corresponding rotational states. $w\left(j\right)$ is 1 for para-$H_2$ (even $j$ states) and 3 for ortho-$H_2$ (odd $j$ states). $D_2O$ is an asymmetric top with its rotational state labeled as $J_{K_a K_c}$ and $v_3$ being the quantum number of antisymmetric stretch, for which the ortho and para states take the odd and even values of the sum ($K_a + K_c + v_3$) and their degeneracy factors are $w\left(j\right)=2$ and $w\left(j\right)=1$, respectively. The quantum internal energy (QIE) at a given temperature is also calculated in the same way by the thermal average over possible energy levels according to their Boltzmann weights,

$$\left\langle E_Q \right\rangle = \sum_{v, j} F_B\left(v, j; T_g\right) E_{vj} \,. \tag{16}$$

It should be noted that the conventional PIMD theory assumes all atoms as distinguishable and cannot take the nuclear spin statistics of ortho/para $H_2$ and $D_2O$ into account. This assumption corresponds to an identical factor $w\left(j\right)$ to all states, or equivalently removing this factor from Eq. (15). We will discuss the effects of neglecting nuclear spin statistics in calculating QIEs and thermally averaged $S_0$ values below.

## II. Additional results

We used the standard procedure to overcome the non-ergodicity of PIMD, which was first proposed by Pérez et al.[18]. As discussed in a recent review[11], because of the frequencies between beads of the ring-polymer depend on $P$, these springs between neighboring beads would become rigid with the increasing P, causing the non-ergodicity[28]. One way to overcome this difficulty is to run a set of short NVE trajectories initiated from uncorrelated snapshots sampled from a thermostatted (NVT) PIMD simulation. This procedure not only correctly samples the Boltzmann weight in the correlation function, but also ensures that each new NVE trajectory explores a different region of microcanonical phase space[11]. Although originally proposed for calculating correlation functions, this procedure has been successfully used in recent direct dynamics simulations[15, 29], giving rise to the proper initial conditions for NVE dynamics calculations. To check this ability, we plot in Fig. S3 the initial distributions of $\theta$ and $\varphi$ of $H_2$ of NVE trajectories, extracted from snapshots after the 1 ps equilibrium PIMD, where the position of a hydrogen atom is given by its centroid. It is found that the angular distributions are generally isotropic within statistical errors, indicating that the $H_2$ molecule is randomly sampled with respect to its orientation. We also check the initial positions and momenta in the phase space for each degree of freedom of each H atom (too many, not shown), which fulfill a Gaussian distribution. This is consistent with what needs to be done in QCT calculations before collision.

Any path integral based method becomes increasingly accurate as $P$ increases. Fig. 1 in the main text shows that the PIMD internal energy of the molecule at moderate and high temperatures converges quickly with $P$. However, the discretization error of a path integral calculation is a function of $\beta_P$. We therefore also characterize the convergence of internal energy at a given temperature as a function of $\beta_P$ (or equivalently $PT_g$), as shown in Fig. S4. One immediately realizes that the convergence of PIMD internal energy at low temperature requires much larger number of beads. We further show the convergence behavior of $S_0$ with respect to $P$ in Fig. S5, which is found to be similar to that of internal energy, namely $S_0$ is also converged with $P = 40$ for $H_2$+Cu(111) and $P = 30$ for $D_2O$+Ni(111).

It is well-known that the antisymmetric requirement of the total wavefunction of Fermions results in a 3:1 ratio of ortho (odd $j$) and para (even $j$) states of $H_2$, and a 2:1 ratio of ortho ($K_a$+$K_c$+$v_3$=odd) and para ($K_a$+$K_c$+$v_3$=even) states of $D_2O$. It has been found that there is little need to distinguish the different nuclear spin species, in case of a simulation at high temperature, for example, for liquid $H_2O$ at room temperature[30] or for rate constants of F+$H_2$ and H+$H_2$ reactions above 200 K[31]. In Fig. S6, the QIEs calculated with and without considering this nuclear spin statistics are compared. There are some very minor differences (< 5 meV) at low temperatures ($T_g$ < 200 K) for $H_2$, and the two sets of results become increasingly indistinguishable with the increasing temperature. For $D_2O$, the difference is negligible in the entire temperature range up to 300 K. At higher temperatures, it becomes difficult to assign the ortho and para states of $D_2O$, but we expect no visible difference there. As a result, we indeed show the QIE results without incorporating nuclear spin statistics in Fig. 1.

To converge the $H_2$ QD dissociative sticking probability on Cu(111) at $T_g$ = 300 K, we compute the state-specific $S_0$ up to $H_2(v = 0, j = 4)$. The results as a function of incidence energy are shown in Fig. S7a. It is found that $S_0$ becomes increasingly higher at low incidence energies with the increasing rotational quantum number. As a result, the thermally averaged $S_0$ values are slightly higher than the ground state ones at very low energies, as shown in Fig. 2 in the main text. Again, we see invisible difference between results with and without including the nuclear spin degeneracy factor of ortho/para $H_2$ (3:1), as displayed in Fig. S7b.

Next, we discuss a possible reason for the less satisfactory performance of RPMD at $E_i$ =0.6~1.0 eV. Indeed, similar phenomenon has been observed in recent NE-RPMD calculations for H scattering on graphene[29], where the RPMD calculated sticking probabilities at high incidence energies are also generally lower than classical ones (note that there is no internal state and ZPE for hydrogen atom). These authors offered no explanation for this difference in that work. To overcome the dissociation barrier, the molecule has to approach the surface closely. However, a too close distance to the surface will cause too strong repulsive force that prevents molecular dissociation. In Fig. S8a, we compare the distributions of closest distance between the molecular center and the surface ($Z_{min}$) during QCT and RPMD nonreactive trajectories, at $E_i$=0.8 eV. It is found that the RPMD distribution obtained by taking the centroid of the ring polymer as the position of atomic hydrogen (referred as RPMD-cen), agrees well with the QCT counterpart. It should be noted, however, that the RPMD theory generates multiple replicas of the system and each of them is evolved in terms of the molecule-surface interaction potential. In this sense, we can evaluate $Z_{min}$ from each replica of $H_2$ as well. One can see in the same Figure that the RPMD distribution obtained by the minimum value of $Z_{min}$ among all replicas (referred as RPMD-min), shifts apparently to the shorter distance end. This indicates that some $H_2$ replicas in RPMD simulations could move much closer to the surface exceeding the average distance described by the centroid. These beads may feel a strong repulsive force and drive the whole molecule back to the vacuum, thus decreasing the dissociation probability. This mechanism would be more important near the hollow site, where the molecule would more easily access to the surface. This hypothesis is confirmed in Fig. S8b by comparing the RPMD-min distributions of $Z_{min}$ during trajectories initiated from the circles centered at the hollow and bridge sites, respectively, with a radius of 0.32 Å (sudden approximation works well in this system). The average minimum value of $Z_{min}$ is smaller for the hollow (1.07 Å) than for the bridge site (1.17 Å). This is consistent with the much smaller $S_0$ at the hollow (0.28) than that at the bridge site (0.48) evaluated from these trajectories. By contrast, the QCT calculate $S_0$ are comparable at the two sites (0.54 and 0.65) by the same analysis. As a result, we attribute the underestimated $S_0$ by RPMD to the fact that some replicas access closer to the surface and feel stronger repulsive force than others to hinder the molecular dissociation. This effect should be less important in the case of $D_2O$ because the heavy oxygen atom (and its replicas) would not approach closely to the surface in any way. It should be noted that this is just a reasonable and qualitative interpretation and more factors may be involved in this process.

Finally, taking the first bending state as an example, we clarify in Fig. S9a that how we scale the corresponding $S_0$ curve of with a varying vibrational efficacy $\eta$ defined in the main text. The horizontal shift in translational energy ($\Delta E_i$) between the ground and excited state $S_0$ curves is determined as $\Delta E_i = E_{vj} \times \eta$, where $E_{vj}$ is the rovibrational excitation energy. A Boltzmann average over all possibly populated states at $T_g = 300$ K gives rise to the QD $S_0$ curve at this temperature, which is shown in Fig. S9b as a function of $\eta$, compared to the ground state QD $S_0$ curve and the RPMD $S_0$ values at $T_g = 300$ K. It is found that the thermal contribution due to internal excitation is important and should be included for a fair comparison with RPMD results. In addition, the $S_0$ curve does not change much with an $\eta$ value close to 1, especially at high translational energies. Again, in Fig. S9c, one notes that the $S_0$ curves with and without including the nuclear spin degeneracy factor of ortho/para $D_2O$ (2:1) are indistinguishable. In any case, the agreement between RPMD and QD results at $T_g = 300$ K is excellent. Also shown in Fig. S9b is the estimated QD $S_0$ values at $T_g = 300$ K by adding up the relative difference of QCT ($T_g = 300$ K) and QCT (ground state), namely $S_0$(QD, 300 K) $= S_0$(QD, ground state)$\times S_0$(QCT, 300 K)$/S_0$(QCT, ground state). Interestingly, the two different ways to correct the thermal effect yield more or less the same results. The agreement with the RPMD and QD $S_0$ at $T_g = 300$ K validates the accuracy of RPMD simulations with respect to the inclusion of quantum effects.

Table. S1 Comparison of rovibrational energy levels (in eV) of $H_2$ obtained by EBK and QD calculations.

|      | $v=0, j=0$ | $v=0, j=1$ | $v=0, j=2$ | $v=0, j=3$ | $v=0, j=4$ |
| --- | --- | --- | --- | --- | --- |
| EBK  | 0.268 | 0.284 | 0.313 | 0.357 | 0.415 |
| QD   | 0.269 | 0.283 | 0.313 | 0.357 | 0.415 |

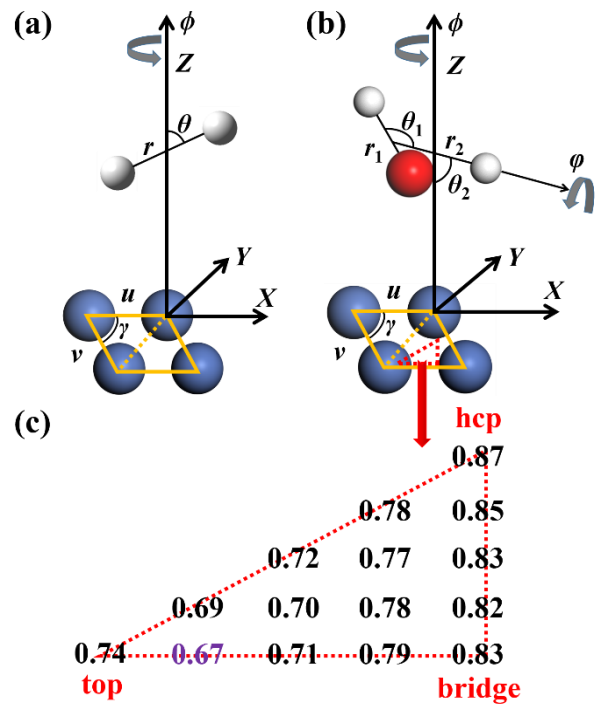

**Fig. S1** Six (a) and nine (b) coordinates defined in our dynamical models for H$_2$+Cu(111) and D$_2$O+Ni(111) systems, respectively. (c) Schematic diagram indicating the 15 site positions and the classical barrier heights in the symmetry unique region of the unit cell used in the site-averaging model. The lowest barrier is marked in purple.

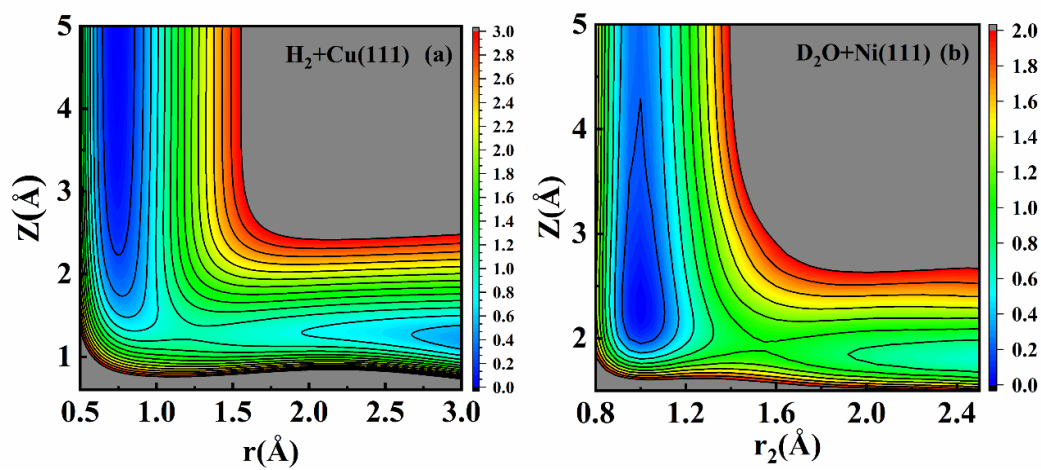

**Fig. S2** Two dimensional contour plots of the potential for the $H_2+Cu(111)$ (a) and $D_2O+Ni(111)$ (b) systems as a function of the molecular height above the surface (Z) and the dissociating coordinate ($r$ or $r_2$). The molecular center is fixed at the bridge site and with $H_2$ parallel to the surface dissociating to two hollow sites for $H_2+Cu(111)$ and the molecular center is fixed at the transition state and with other angular DOFs optimized for $D_2O+Ni(111)$.

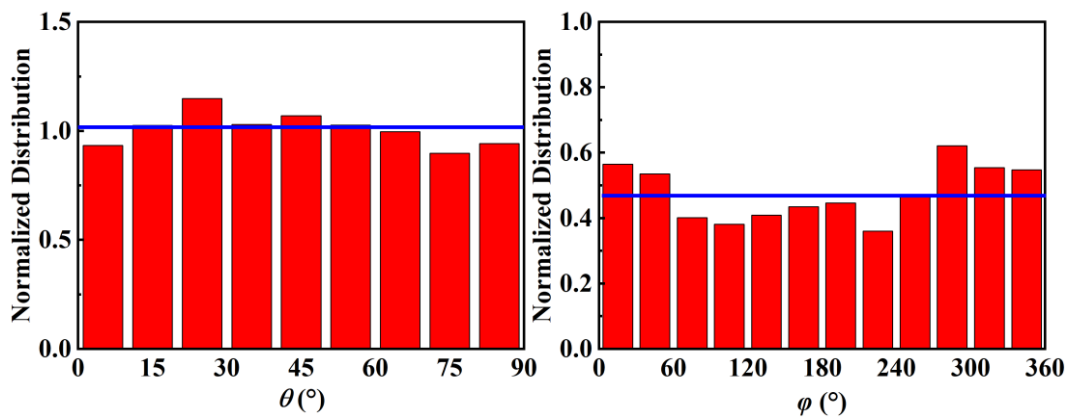

**Fig. S3** Normalized distributions of polar angle P(θ) (a) and azimuth angle P(φ) (b) of initial $H_2$ configurations far from Cu(111) for the NE-RPMD simulations taken from equilibrated PIMD trajectories with the Andersen thermostat at temperature of 300 K. Note that P(θ) for polar angle is divided by the normalization factor sin(θ). The blue line corresponds to the theoretical uniform distribution as a guide of eyes.

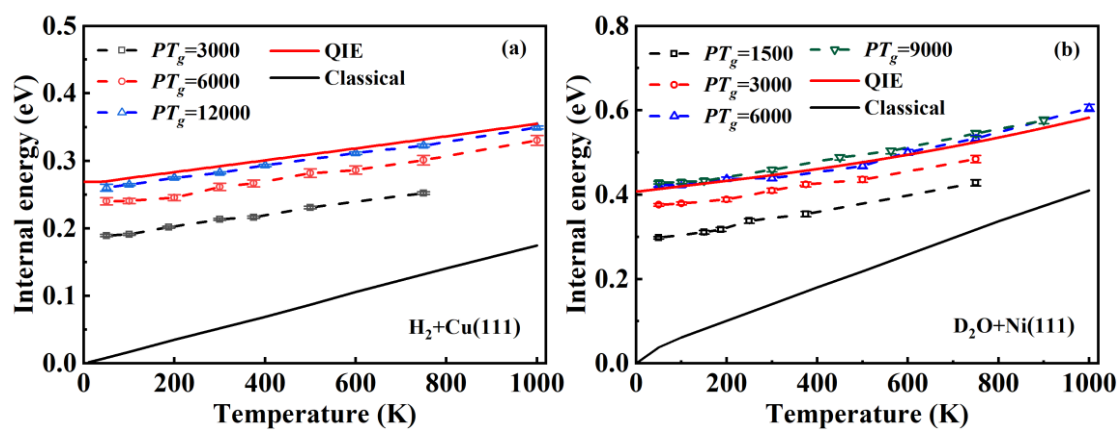

**Fig. S4** PIMD internal energies of $H_2$ (a) and $D_2O$ (b) estimated from NVT simulations of $H_2$+Cu(111) and $D_2O$+Ni(111) systems via virial centroid theorem, as a function of temperature at fixed $\beta_P$ (equivalently $PT_g$) values, compared to classical (black line) and quantum internal energy (red line).

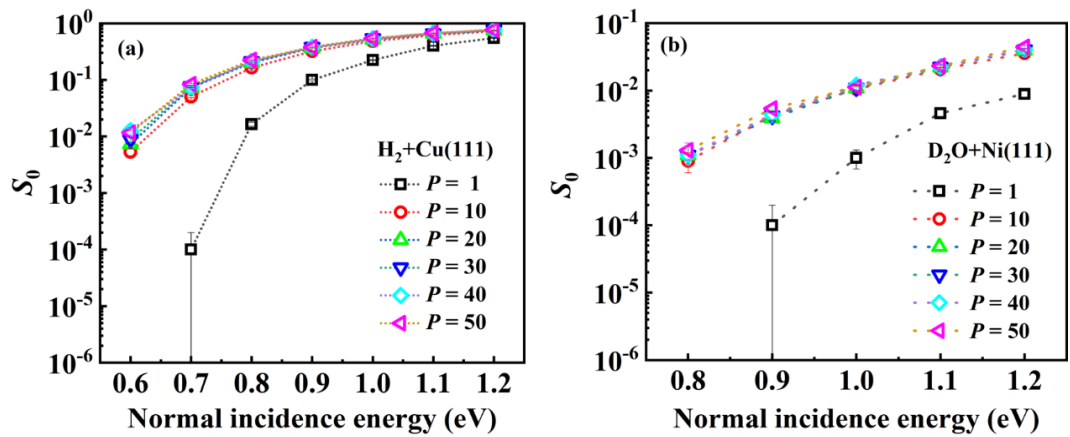

**Fig. S5** Dependence of $S_0$ on the number of beads for $H_2$ on Cu(111) (a) and $D_2O$ on Ni(111) (b), at $T_g$= 300 K.

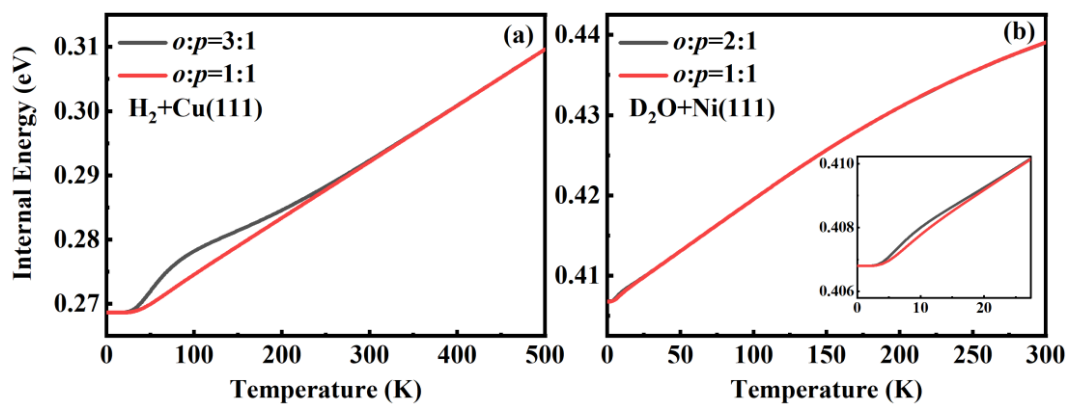

**Fig. S6** Quantum internal energies of $H_2$ (a) and $D_2O$ (b) calculated with (black) and without (red) consideration of nuclear spin statistics (ortho:para=3:1 for $H_2$ and 2:1 for $D_2O$). The inset of panel (b) illustrates the subtle difference of two sets of calculations at very low temperature.

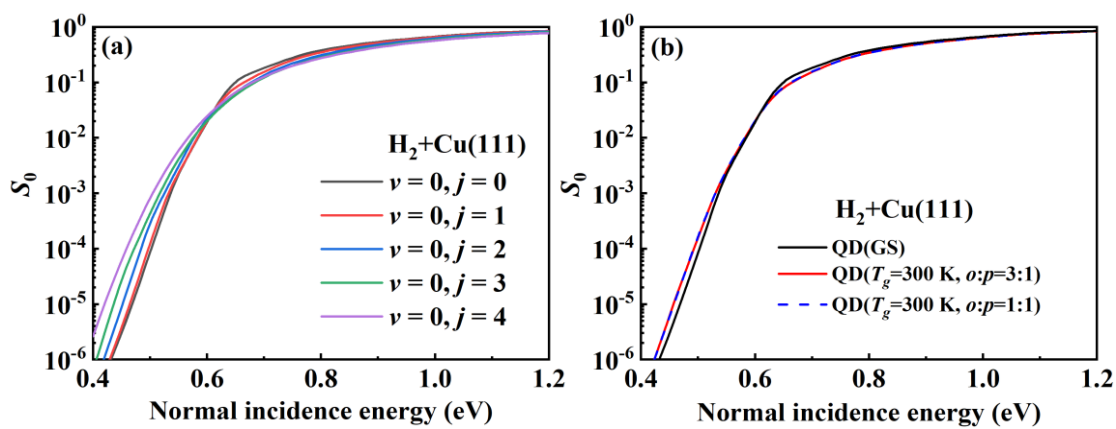

**Fig. S7** (a) State selected dissociation probabilities for H₂($v = 0$, $j = 0$-4) on Cu(111). (b) thermally averaged dissociation probabilities at $T_g = 300$ K calculated with (red solid) and without (blue dash) consideration of nuclear spin statistics (ortho:para=3:1 for H₂), compared with the ground state reactivity (black solid) of H₂ on Cu(111).

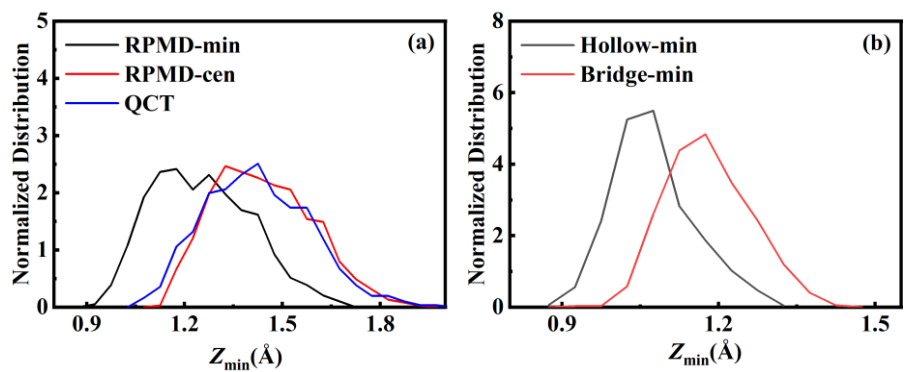

**Fig. S8** (a) Distributions of the closest vertical distance between $H_2$ and the Cu(111) surface ($Z_{min}$) during scattered trajectories obtained from QCT (blue), RPMD with the center of centroid (RPMD-cen, red), and RPMD with the minimum value of $Z_{min}$ among all replicas (RPMD-min, black). (b) Comparison of RPMD-min distributions obtained with trajectories initiated in a region centered at the bridge site (red) and hollow site (black), respectively, with a radius of 0.32 Å.

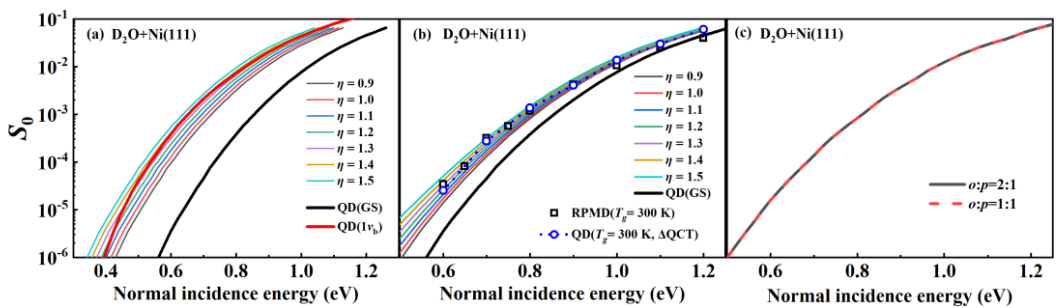

**Fig. S9** (a) Approximate $S_0$ curves (thin lines) of the first bending excited state of $D_2O$ ($1v_b$) shifted from the baseline of ground state (thick black line) by a varying vibrational efficacy ($\eta$), compared to the exact calculated $S_0$ curve of $D_2O$ ($1v_b$) (thick red line). (b) Boltzmann average of QD $S_0$ curves over all possibly populated states at $T_g$=300 K with different universal $\eta$ (thin lines). The QD $S_0$ curve for the ground state (thick black line), RPMD results (black squares), and QD results estimated from QCT at $T_g$=300 K (blue circle), are shown for comparison. (c) Boltzmann average of QD $S_0$ curves over all possibly populated states at $T_g$=300 K, obtained with (black line) and without (red dash line) considering nuclear spin statistics (ortho:para= 2:1 for $D_2O$).

# References:

1.  Dai, J.; Zhang, J. Z. H. Quantum adsorption dynamics of a diatomic molecule on surface: Four-dimensional fixed-site model for $H_2$ on Cu(111). *J. Chem. Phys.* **1995**, *102* (15), 6280-6289.

2.  Perdew, J. P.; Wang, Y. Accurate and simple analytic representation of the electron-gas correlation energy. *Phys. Rev. B* **1992**, *45* (23), 13244–13249.

3.  Jiang, B.; Guo, H. Dynamics of water dissociative chemisorption on Ni(111): Effects of impact sites and incident angles. *Phys. Rev. Lett.* **2015**, *114*, 166101.

4.  Jiang, B.; Guo, H. Permutation invariant polynomial neural network approach to fitting potential energy surfaces. III. Molecule-surface interactions. *J. Chem. Phys.* **2014**, *141*, 034109.

5.  Hu, X.; Hase, W. L.; Pirraglia, T. Vectorization of the general Monte Carlo classical trajectory program VENUS. *J. Comput. Chem.* **1991**, *12*, 1014-1024

6.  Gutzwiller, M. C. *Chaos in Classical and Quantum Mechanics*. Springer: New York, 1990.

7.  Porter, R. N.; Raff, L. M.; Miller, W. H. Quasiclassical selection of initial coordinates and momenta for a rotating Morse oscillator. *J. Chem. Phys.* **1975**, *63*, 2214.

8.  Peslherbe, G. H.; Wang, H.; Hase, W. L. Monte Carlo sampling for classical trajectory simulations, a chapter in Monte Carlo methods in chemical physics. *Adv. Chem. Phys.* **1999**, *105*, 171-201.

9.  Chandler, D.; Wolynes, P. G. Exploiting the isomorphism between quantum theory and classical statistical mechanics of polyatomic fluids. *J. Chem. Phys.* **1981**, *74*, 4078-4095.

10. Craig, I. R.; Manolopoulos, D. E. Quantum statistics and classical mechanics: Real time correlation frunction from ring polymer molecular dynamics. *J. Chem. Phys.* **2004**, *121*, 3368-3373.

11. Habershon, S.; Manolopoulos, D. E.; Markland, T. E.; Miller III, T. F. Ring-polymer molecular dynamics: Quantum effects in chemical dynamics from classical trajectories in a extended phase space. *Annu. Rev. Phys. Chem.* **2013**, *64*, 387-413.

12. Welsch, R.; Song, K.; Shi, Q.; Althorpe, S. C.; Miller, T. F. Non-equilibrium dynamics from RPMD and CMD. *J. Chem. Phys.* **2016**, *145* (20), 204118.

13. Habershon, S.; Manolopoulos, D. E. Zero point energy leakage in condensed phase dynamics: an assessment of quantum simulation methods for liquid water. *J. Chem. Phys.* **2009**, *131* (24), 244518.

14. Herman, M. F.; Bruskin, E. J.; Berne, B. J. On path integral Monte Carlo simulations. *J. Chem. Phys.* **1982**, *76* (10), 5150-5155.

15. Suleimanov, Y. V.; Aguado, A.; Gómez-Carrasco, S.; Roncero, O. A ring polymer molecular dynamics approach to study the transition between statistical and direct mechanisms in the $H_2 + H_3^+ \rightarrow H_3^+ + H_2$ reaction. *J. Phys. Chem. Lett.* **2018**, *9*, 2133-2137.

16. Andersen, H. C. Molecular dynamics simulations at constant pressure and/or temperature. *J. Chem. Phys.* **1980**, *72*, 2384-2393.

17. Suleimanov, Y. V.; Aoiz, F. J.; Guo, H. Chemical reaction rate coefficients from ring polymer molecular dynamics: Theory and practical applications. *J. Phys. Chem. A* **2016**, *120* (43), 8488-8502.

18. Pérez, A.; Tuckerman, M. E.; Müser, M. H. A comparative study of the centroid and ring-polymer molecular dynamics methods for approximating quantum time correlation functions from path integrals. *J. Chem. Phys.* **2009**, *130* (18), 184105.

19. Jiang, B.; Guo, H. Six-dimensional quantum dynamics for dissociative chemisorption of $H_2$ and $D_2$ on Ag(111) on a permutation invariant potential energy surface. *Phys. Chem. Chem. Phys.* **2014**, *16*, 24704-24715.

20. Jiang, B.; Song, H.; Yang, M.; Guo, H. Quantum dynamics of water dissociative chemisorption on rigid Ni(111): An approximate nine-dimensional treatment. *The Journal of Chemical Physics* **2016**, *144*

(16), 164706.

21. Liu, T.; Fu, B.; Zhang, D. H. Six-dimensional quantum dynamics study for the dissociative adsorption of HCl on Au(111) surface. *J. Chem. Phys.* **2013**, *139*, 184705.

22. Liu, T.; Fu, B.; Zhang, D. H. Validity of the site-averaging approximation for modeling the dissociative chemisorption of $H_2$ on Cu(111) surface: A quantum dynamics study on two potential energy surfaces. *J. Chem. Phys.* **2014**, *141*, 194302.

23. Hu, X.; Jiang, B.; Xie, D.; Guo, H. Site-specific dissociation dynamics of $H_2/D_2$ on Ag(111) and Co(0001) and the validity of the site-averaging model. *J. Chem. Phys.* **2015**, *143*, 114706.

24. Zhang, Z.; Liu, T.; Fu, B.; Yang, X.; Zhang, D. H. First-principles quantum dynamical theory for the dissociative chemisorption of $H_2O$ on rigid Cu(111). *Nat. Comm.* **2016**, *7*, 11953.

25. Jiang, B.; Guo, H. Quantum and classical dynamics of water dissociation on Ni(111): A test of the site-averaging model in dissociative chemisorption of polyatomic molecules. *J. Chem. Phys.* **2015**, *143* (16), 164705.

26. Jackson, B.; Nave, S. The dissociative chemisorption of methane on Ni(100): Reaction path description of mode-selective chemistry. *J. Chem. Phys.* **2011**, *135*, 114701.

27. Diaz, C.; Olsen, R. A.; Busnengo, H. F.; Kroes, G.-J. Dynamics on six-dimensional potential energy surfaces for $H_2$/Cu(111): Corrugation reducing procedure versus modified Shepard interpolation method and PW91 versus RPBE. *J. Phys. Chem. C* **2010**, *114* (25), 11192-11201.

28. Hall, R. W.; Berne, B. J. Nonergodicity in path integral molecular dynamics. *J. Chem. Phys.* **1984**, *81* (8), 3641-3643.

29. Jiang, H.; Kammler, M.; Ding, F.; Dorenkamp, Y.; Manby, F. R.; Wodtke, A. M.; Miller, T. F.; Kandratsenka, A.; Bünermann, O. Imaging covalent bond formation by H atom scattering from graphene. *Science* **2019**, *364* (6438), 379.

30. Vega, C.; Abascal, J. L. F. Simulating water with rigid non-polarizable models: a general perspective. *Phys. Chem. Chem. Phys.* **2011**, *13* (44), 19663-19688.

31. Collepardo-Guevara, R.; Suleimanov, Y. V.; Manolopoulos, D. E. Bimolecular reaction rates from ring polymer molecular dynamics. *J. Chem. Phys.* **2009**, *130*, 174713.